\documentclass{article}

\usepackage[preprint]{neurips_2023}
\usepackage{graphicx}
\usepackage{longtable}
\usepackage{tcolorbox}
\usepackage{amsmath} 
\usepackage{multirow}
\usepackage{xcolor}
\usepackage{array}
\usepackage{hyperref}
\hypersetup{
    colorlinks=true,
    urlcolor=blue!60!black
}

\usepackage[utf8]{inputenc} 
\usepackage[T1]{fontenc}    
\usepackage{hyperref}       
\usepackage{url}            
\usepackage{booktabs}       
\usepackage{amsfonts}       
\usepackage{nicefrac}       
\usepackage{microtype}      
\usepackage{xcolor}         
\usepackage{enumitem}
\usepackage{parskip}  
\title{Audio-Reasoner: Improving Reasoning Capability in Large Audio Language Models}

\author{%
  Zhifei Xie$^{1\dag}$ \quad Mingbao Lin$^{3\dag}$ \quad Zihang Liu$^{2\dag}$ 
  \\ 
  \textbf{Pengcheng Wu}$^{1}$ \quad \textbf{Shuicheng Yan}$^{2\ddag}$ \quad \textbf{Chunyan Miao}$^{1\ddag}$
  \\
  $^1$Nanyang Technological University    
  $^2$National University of Singapore  
  $^3$Rakuten  \\
  $^{\dag}${\small Equal Contributions}  \; $^{\ddag}$Corresponding Authors \\
  \texttt{\small zhifei001@e.ntu.edu.sg  linmb001@outlook.com  liuzihang99@gmail.com} \\
  \texttt{\small pengchengwu@ntu.edu.sg   yansc@nus.edu.sg  ascymiao@ntu.edu.sg} \\
  {Project}: \url{https://github.com/xzf-thu/Audio-Reasoner}
}
\begin{document}
\maketitle




\begin{abstract}
Recent advancements in multimodal reasoning have largely overlooked the audio modality. We introduce Audio-Reasoner, a large-scale audio language model for deep reasoning in audio tasks. We curate a diverse collection of multi-task audio datasets with simple annotations, refining them through structured secondary labeling and complex question generation. Additionally, We utilize advanced closed-source models to generate structured reasoning chains, transforming raw annotations into a formatted inference process. Following inference scaling principles, we train Audio-Reasoner on CoTA, a high-quality reasoning dataset with 1.2 million reasoning-rich samples. Experiments show state-of-the-art performance across key benchmarks, including MMAU-mini (+25.42\%), AIR-Bench chat/foundation(+14.57\%/+10.13\%), and MELD (+8.01\%). Our findings stress the core of structured CoT training in advancing audio reasoning. 
\end{abstract}

\begin{figure}[!h]
    \centering
    \includegraphics[width=0.98\linewidth]{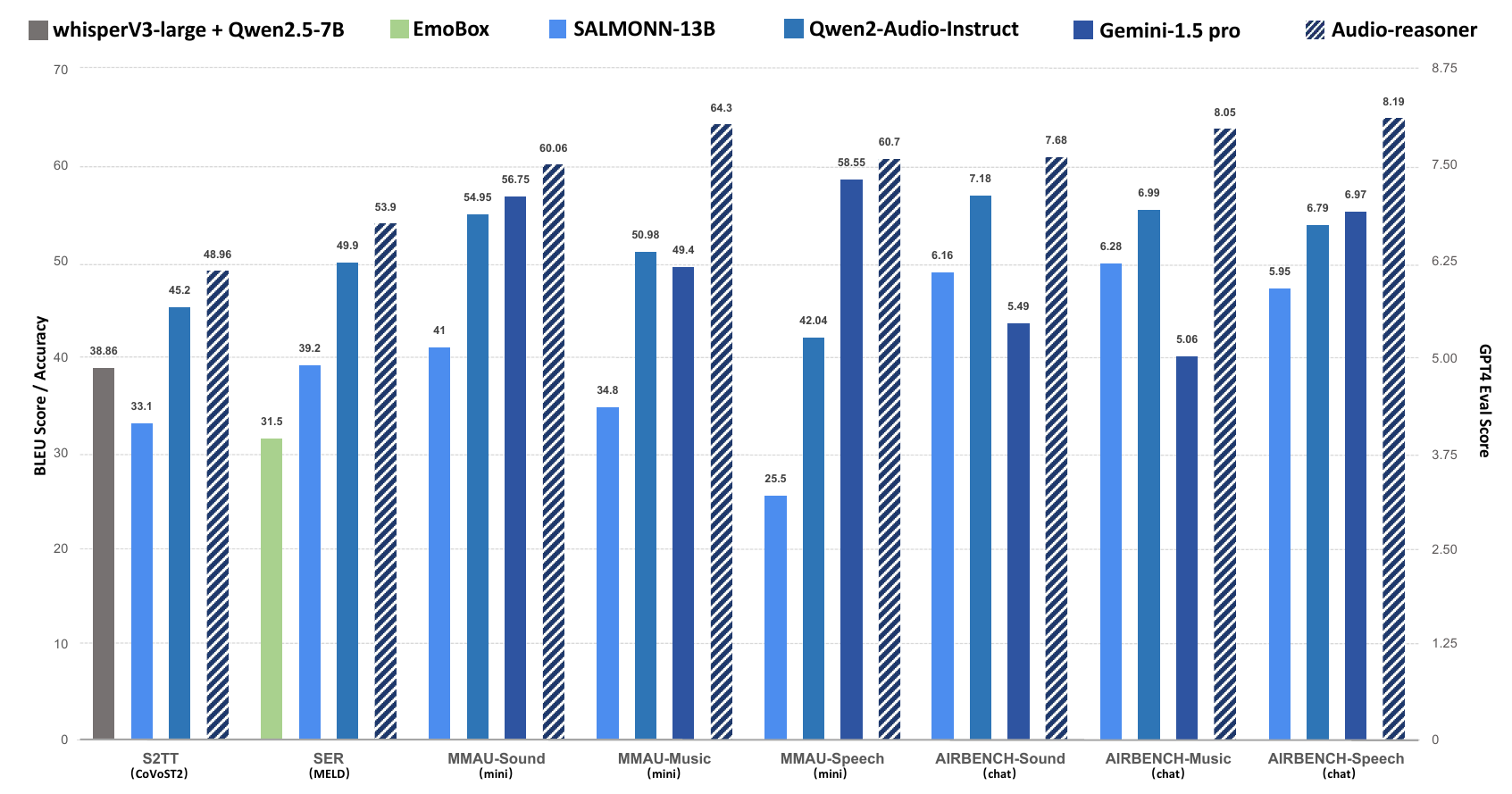}
    \caption{Benchmark performance of Audio-Reasoner on tasks of speech emotion recognition (SER), speech-to-text translation (S2TT), MMAU, and AIR-Bench chat. S2TT is measured using BLEU while SER and MMAU are measured ACC (left axis). AIR-Bench is scored by GPT (right axis).}
    \label{fig:performance_benchmark}
\end{figure}

\section{Introduction}
Recent advancements in large language models (LLMs) have significantly enhanced their reasoning capabilities, driven by innovations like chain-of-thought (CoT) and inference scaling.
Models such as OpenAI's o1~\citep{o1} and Deepseek-R1~\citep{r1} have demonstrated significant improvements, especially in tasks like mathematics and coding~\citep{kimi1.5,marco,s1,deepseekv3,o1coder,deng2024explicit,hui2024qwen2code,qwen2math}. 
These models have set a precedent for achieving ``deep thinking'' by tackling complex, structured tasks. The CoT framework, in particular, has been successfully applied to multimodal models, improving their cognitive abilities in image and video reasoning. Models such as Visual-CoT~\citep{visualcot}, LLaVA-Reasoner~\citep{llavareasoner}, and MAmmoTH-VL~\citep{mammoth} have demonstrated that large-scale datasets and mdultimodal reasoning can enhance model performance, even in tasks involving images and videos~\citep{zou2023generalizable}. Other works, like Mulberry~\citep{mulberry} and Image-of-Thought~\citep{imageofthought}, integrate reflective and image-editing tools to further refine multimodal reasoning, offering new pathways for understanding complex queries across multiple modalities.

However, the application of CoT in the audio modality has been largely underexplored. Although models like Audio Flamingo~\citep{audioflamingo}, SALMONN~\citep{salmonn}, and Qwen2-Audio~\citep{qwen2audio} have pushed the boundaries of large audio language models (LALMs), these advancements have not yet incorporated CoT reasoning at scale. Recent research~\citep{audiocot} suggests that existing CoT methods, such as zero-shot reasoning in audio tasks, fail to significantly improve performance on more complex queries. This limitation is largely attributed to the simplicity of existing audio datasets---such as AudioSet~\citep{audioset}, AudioCaps~\citep{audiocaps}, and Clotho~\citep{clotho}---which predominantly feature short, simple labels. These simplified datasets hinder the development of LALMs capable of more intricate reasoning. Without richer, more complex data, these models struggle with long-form reasoning, and the application of CoT often leads to severe hallucinations and degraded performance. Therefore, advancing CoT in LALMs necessitates overcoming these dataset limitations to allow for more effective, deep reasoning.

To address the challenges in audio-based reasoning, we propose a scalable and effective approach to generating high-quality pretraining data. Using state-of-the-art commercial models, we introduce CoTA, a large-scale dataset containing \textbf{1.2 million} refined captions and question-answer (QA) pairs. CoTA spans multiple datasets and tasks, undergoing rigorous filtering to ensure diversity and quality.
Building on CoTA, we develop Audio-Reasoner, a large audio language model designed for long-context reasoning. Audio-Reasoner is trained with a 4K token context window and generates structured CoT reasoning with length could more than exceeding 1K tokens in real-world tasks. The model is fine-tuned on CoTA using supervised fine-tuning, adhering to a structured reasoning framework, as illustrated in Figure\,\ref{fig:inference_plp}:
(1) Planning---Identifies key problem components from the user query and formulates a structured sequence of reasoning steps essential for deriving an answer.
(2) Caption---Extracts and integrates relevant multimodal content from the input to enrich the reasoning process.
(3) Reasoning--- Executes a systematic, step-by-step reasoning procedure to ensure logical coherence and accuracy.
(4) Summary---Synthesizes the reasoning process into a final response that is concise, contextually grounded, and precise.

Our experimental results, partially presented in Figure\,\ref{fig:performance_benchmark}, demonstrate the effectiveness of Audio-Reasoner. More comprehensively, we evaluate the model across multiple benchmarks:
MMAU-mini~\citep{mmau}: A dataset with 1,500 closed-choice questions testing reasoning across sound, speech, and music.
AIR-Bench~\citep{air}: Various types of audio signals including human speech, natural sounds, and music.
CoVoST2(zh-en)~\citep{covost}: Speech-to-text translation in Chinese and English.
MELD~\citep{meld}: Emotion classification.
Compared to Qwen2-Audio-Instruct~\citep{qwen2audio}, Audio-Reasoner achieves:
+25.4\% improvement on MMAU-mini with reasoning subtask gains: +44.4\%, +26.1\%, and +9.3\%; 
+14.6\% gains on AIR-Bench chat;
+30.6\% on CoVoST2(ZN/EN subset, Average BLEU score.); 
+8.01\% on MELD.
These results validate the effectiveness of our approach in advancing long-context reasoning and inference scaling for audio models.

The major contributions we have made in this paper include:
\begin{itemize}[leftmargin=*]
\item We propose Audio-Reasoner, designed for deep reasoning and inference scaling in the audio modality. Built upon Qwen2-Audio and fine-tuned with structured CoT training, Audio-Reasoner significantly improves long-context reasoning across diverse audio tasks.
\item We develop CoTA, a large-scale dataset with 1.2 million high-quality captions and QA pairs, spanning multiple audio domains. The dataset enables structured reasoning and enhances audio-language pretraining.
\item We introduce a scalable data generation pipeline leveraging advanced commercial models to produce complex reasoning-based QA pairs and structured CoT annotations, enriching model training.
\item We achieve state-of-the-art performance, with +25.4\% gains over Qwen2-Audio-Instruct on MMAU-mini, along with significant improvements in reasoning, translation, and emotion recognition tasks.
\end{itemize}

\section{Related Work}

\textbf{Chain-of-Thought Reasoning}. LLMs leverage in-context learning (ICL) to enhance their reasoning capabilities by processing prompts and context. This is further strengthened through CoT techniques. Various CoT methods have been explored, including Tree of Thoughts (TOT)~\citep{tree}, manual few-shot CoT~\citep{mannual}, and  automatically generated approaches~\citep{auto1,auto2}. In addition, studies have delved into the necessity of CoT, its theoretical foundations, and its effectiveness across a wide range of tasks~\citep{whycot1s,whycot2,whycot3}. The release of OpenAI's o1 model~\citep{o1} has sparked renewed interest in CoT research, significantly boosting the capabilities of LLMs, especially in multi-step reasoning tasks such as coding~\citep{o1coder} and mathematics~\citep{qwen2math}, setting new performance benchmarks. CoT techniques have been integrated with other methods such as Monte Carlo Tree Search~\citep{monte}, reflection~\citep{r1}, and tool use~\citep{toolllm}, and are often trained alongside reinforcement learning approaches~\citep{DPO,KPO,TRPO,deepseekmath}.

\textbf{Multimodal Chain-of-Thought}. CoT techniques have also been explored in the realm of multimodal large models. For example, Visual-COT~\citep{visualcot} incorporates object detection to aid in reasoning, LLaVA-Reasoner~\citep{llavareasoner} uses closed-source models for CoT fine-tuning through recaptioning. LLaVA-CoT~\citep{llavacot} and MAmmoTH-VL~\citep{mammoth} scale datasets to improve model performance. Other models, such as Mulberry~\citep{mulberry}, explore application of reflective thinking, and Image-of-Thought~\citep{imageofthought} integrates image editing tools to enhance reasoning. Video-related studies~\citep{videocot,videoespresso,fei2024video,tang2024cardiff} have demonstrated the effectiveness of CoT in reasoning tasks within the video domain. However, the application of CoT in the audio domain is still in its infancy. The study Audio-COT~\citep{audiocot} shows some improvement with zero-shot COT in audio tasks, but it falls short on more complex problems. This paper aims to explore this gap further.

\textbf{Large Audio Language Models}. LALMs can be broadly categorized into two areas: audio understanding and real-time dialogue. Audio understanding models typically consists of a three-layer architecture---an encoder, connector, and an LLM---focusing on specific domains, as seen in models like Mu-LLaMA~\citep{mullama}, LTU~\citep{LTU}, EmoBox~\citep{emobox}, and GAMA~\citep{gama}. Other models, such as LTU-AS~\citep{LTUas}, SALMONN~\citep{salmonn} and Qwen2-Audio~\citep{qwen2audio}, employ unified architectures designed for multi-task training. Real-time conversation models, which focus on speech input and extend transformers to real-time speech synthesis, are also gaining popularity~\citep{zhang2023speechgpt,xie2024mini,xie2024mini2,fu2025vita,defossez2024moshi}. However, despite their focus on understanding and rapid response, current LALMs still lack significant exploration into reasoning tasks, such as COT. This paper addresses this gap by investigating the application of CoT in LALMs.

\section{Audio-Reasoner}
In this section, we present the training methodology for our Audio-Reasoner model, designed to enable LALMs with structured, step-by-step reasoning in multimodal tasks. Unlike standard models, Audio-Reasoner generates both a chain-of-thought (CoT) and final response, ensuring logical consistency and interpretability. We describe the training framework, which includes a structured CoT approach, a novel data generation pipeline, and the large-scale CoTA dataset. This dataset is carefully constructed through automated annotation, reasoning chain creation, and validation, setting the foundation for robust and explainable audio-based reasoning.

\begin{figure}[!t]
    \centering
    \newpage
    \includegraphics[width=1.0\linewidth]{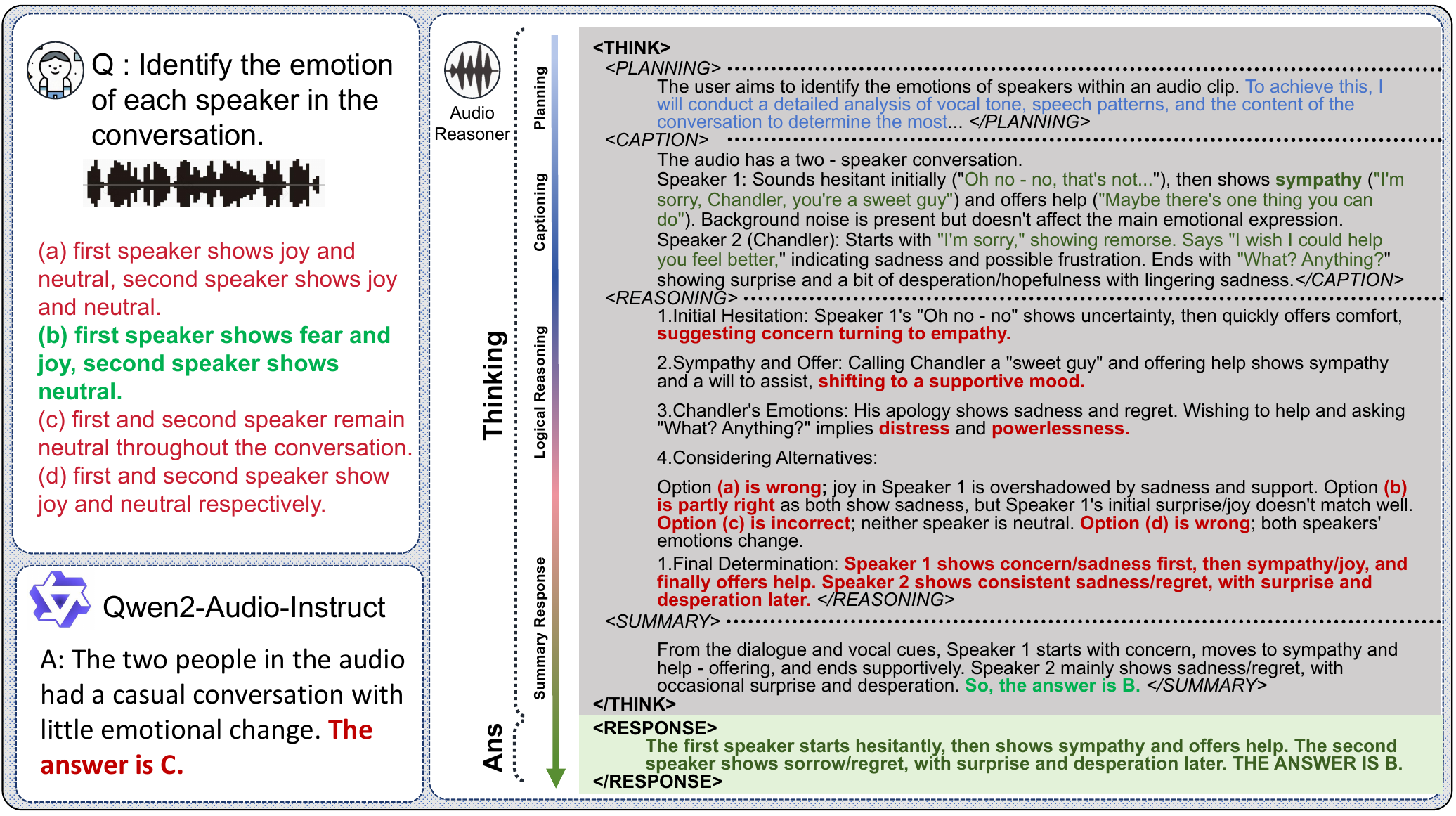}
    \caption{Comparison between Audio-Reasoner and Qwen2-Audio-Instruct: While Qwen2-Audio-Instruct produces brief and error-prone reasoning, our Audio-Reasoner uses a structured reasoning framework with distinct ``thinking'' and ``response'' phases, ensuring a more reliable and well-supported output through careful planning, information extraction, and step-by-step reasoning.}
    \label{fig:inference_plp}
\end{figure}

\subsection{Model Training with Audio Reasoning}
A standard large language model is trained to generate an output sequence $Y$ given an input sequence $X$. The probability distribution of the model's output is formulated as:
\begin{equation}
    P(Y|X;\theta) = f_{\theta}(X),
\end{equation}
where $f_\theta$ is a Transformer-based model parameterized by $\theta$. The training objective follows a maximum likelihood estimation framework:
\begin{equation}
    \mathcal{L}(\theta) = -\sum_{i=1}^N\log P(Y_i|X_i;\theta).
\end{equation}

In our Audio-Reasoner, the input consists of an audio signal $A$ and a text-based query $Q$, forming the multimodal input representation:
\begin{equation}
    X = (A, Q).
\end{equation}

Unlike conventional LLMs, where the output is a single response, we structure the model's output into two distinct components: the chain of thought reasoning $C$, which captures the step-by-step logical process, and the final response $R$, which provides the ultimate answer. The model thus learns to generate the concatenation of $C$ and $R$, leading to the probability distribution:
\begin{equation}
    P(C, R | A, Q; \theta) = f_\theta(A, Q).
\end{equation}

To ensure explicit learning of both reasoning and final response generation, we construct a dataset defined as:
\begin{equation}
    \mathcal{D} = \{(A_i, Q_i, C_i, R_i)\}_{i=1}^N,
\end{equation}
where each training sample consists of an input audio signal $A_i$, its corresponding textual query $Q_i$, the structured reasoning process $C_i$, and the final answer $R_i$. This dataset formulation reinforces the model's ability to perform in-context learning and deep reasoning, ensuring that generated responses are not only accurate but also logically structured.

The training objective maximizes the likelihood of both $C$ and $R$, encouraging the model to first reason and then generate a response. The loss function is given by:
\begin{equation}
    \mathcal{L}(\theta) = -\sum_{i=1}^N \log P(C_i, R_i | A_i, Q_i; \theta).
\end{equation}

By optimizing this objective, Audio-Reasoner is trained to articulate a structured reasoning process before providing its final response. This approach enhances interpretability, reliability, and alignment with human reasoning.

At inference-time, our Audio-Reasoner follows a structured reasoning pipeline, as illustrated in Figure\,\ref{fig:inference_plp}. The reasoning process consists of four sequential steps: 
(1) \textbf{Planning} ($P$): The model analyzes the query, identifies key problem components, and outlines the reasoning steps necessary to derive an answer.
(2) \textbf{Captioning} ($C$): Relevant multimodal content is extracted from the input, such as speech transcription, acoustic event detection, or context information.
(3) \textbf{Reasoning} ($R$): Based on the extracted content, the model performs structured, step-by-step reasoning.
(4) \textbf{Summary} ($S$): The model synthesizes its reasoning process into a final, concise, and accurate response.
This structured inference process can be formalized as follows:
\begin{align}
P &\sim f_{\theta}(A, Q), \\
C &\sim f_{\theta}(A, Q, P), \\
R &\sim f_{\theta}(A, Q, P, C), \\
S &\sim f_{\theta}(A, Q, P, C, R).
\end{align}

Compared to the direct-response counterpart~\citep{qwen2audio}, this approach provides two key advantages:
\textbf{Improved Interpretability}---By explicitly modeling each reasoning step, the process becomes more transparent, making it easier to analyze and diagnose errors.
\textbf{Reduced Hallucinations}---The structured reasoning pipeline mitigates speculative or incorrect responses, ensuring that outputs remain logically grounded.

Figure\,\ref{fig:inference_plp} illustrates the structured CoT reasoning process, highlighting how each stage contributes to the final response. This approach draws inspiration from recent advancements in symbolic reasoning and CoT training~\citep{theoretical}, which emphasize that zero-shot reasoning without training is less effective. Moreover, previous studies have shown that models tuned on native CoT data significantly outperform those trained on generic labels, especially in multimodal reasoning tasks~\citep{mammoth,mulberry}.

\begin{table}[ht]
    \centering
    \caption{Domains and tasks of our constructed CoTA dataset.}
    \begin{tabular}{ccc}
        \toprule
        \textbf{Domain} & \textbf{Task} & \textbf{Description} \\
        \midrule
        Sound & Sound QA & Sound question answering \\
        \midrule
        \multirow{3}{*}{Speech} & Speech QA & Speech question answering \\
         & SER & Speaker emotion recognition \\
         & S2TT & Speech to text translation \\
        \midrule
        Music & Music QA & Music question answering \\
        \bottomrule
    \end{tabular}
    \label{tab:description}
\end{table}

\subsection{Systematic Data Preparation for Audio Reasoning}
Training the Audio-Reasoner model requires a high-quality, diverse, and multitask audio-based reasoning dataset. Our goal is to develop a scalable and effective data generation method that systematically transforms raw audio data and simple human-labeled annotations into structured reasoning tasks. The resulting CoTA dataset with 1.2 million samples, focusing on  complex reasoning-based question-answering tasks, spans three domains---audio, speech, and music---as detailed in Table\,\ref{tab:description}.

\begin{figure}[!t]
    \centering
    \includegraphics[width=1.0\linewidth]{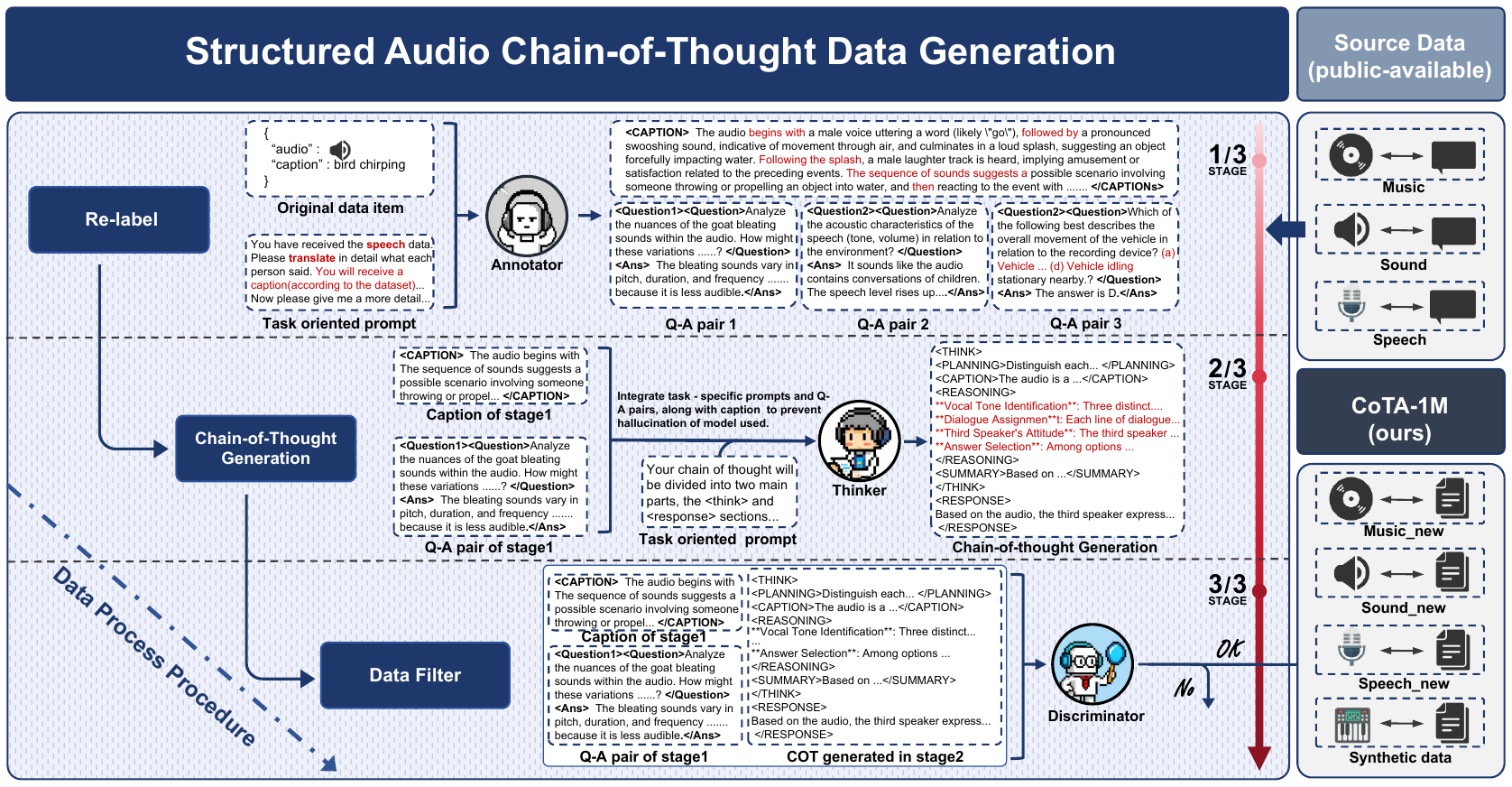 }
    \caption{Multistage data generation pipeline.}
    \label{fig:pipeline}
\end{figure}

To achieve this, we design a structured data generation pipeline consisting of three key stages: (1) generating high-quality annotations and diverse questions, (2) constructing structured reasoning chains, and (3) performing comprehensive validation. The complete pipeline is illustrated in Figure\,\ref{fig:pipeline}. The following sections describe each stage in detail.

\subsubsection{Multistage Data Generation Pipeline}
\textbf{Stage 1: Automated Annotation and Question-Answer Synthesis}.We begin by using advanced external closed-source models to improve the existing simple human annotations into high-quality and logically coherent training data. While large language models can produce hallucinations in free-form generation, they are highly effective at structured, evidence-based content creation. We leverage this strength by instructing the model to sequentially describe the audio elements, enhancing its understanding of sound sources and speech context. Based on these enriched descriptions, the model generates a diverse set of questions, ranging from straightforward factual queries to complex logical reasoning problems.  This process ensures that the dataset captures a broad spectrum of reasoning tasks, as summarized in Table\,\ref{tab:description}. Besides, in Sec.\,\ref{sec:prompt_stage1} we provide the prompt in deducing closed-source models to generate questions and corresponding answers.

\textbf{Stage 2: Structured Reasoning Chain Construction}. Next, we transform the generated question-answer pairs into structured reasoning chains. Given the limited development of CoT methodologies in the audio domain, we adopt a systematic approach to ensure inference stability. The model first plans and analyzes the questions, extracts key information from the captions, and formulates logical steps leading to the answer. To facilitate structured reasoning, we employ explicit step demarcations such as <THINK> and <REASONING>, allowing the model to autonomously construct multi-step inference pathways. Sec.\,\ref{sec:prompt_stage2} describes the prompt used for the structured reasoning chain construction process. 

\textbf{Stage 3: Quality Assurance and Dataset Validation}. Finally, we subject the generated data to a rigorous review process. Using the raw audio input, Stage 1 annotations, and Stage 2 reasoning chains, the model assesses whether the generated content is accurate, coherent, and suitable for inclusion in the final dataset. This step ensures the overall quality and reliability of the CoTA dataset. Sec.\,\ref{sec:prompt_stage3} illustrates prompt used for filtering low-quality contents.

\subsubsection{Task Taxonomy: Categories and Reasoning Methodologies}

The CoTA dataset encompasses a range of reasoning-based tasks, each requiring distinct reasoning path that the model should grasp. These include:

\textbf{(1) Sound-Based Question Answering}: The model identifies and analyzes sound characteristics, contextualizing them within the user's query to derive a reasoned response.
\textbf{(2) Speech-Based Question Answering}: The model recognizes speaker timbres, transcribes speech content, and incrementally processes the question to determine the appropriate answer.
\textbf{(3) Speech Emotion Recognition (SER) and Speech-to-Text Translation (S2TT)}: These specialized tasks require the model to integrate speech recognition with emotion analysis and language translation, forming a structured reasoning process.
\textbf{(4) Music-Based Question Answering}: As music is highly abstract, the model first analyzes fundamental attributes such as tonality, tempo, and emotion before progressing to genre classification and deeper inferential reasoning based on the user's query.
A comprehensive breakdown of the dataset's task categories and reasoning pathways is provided in Table\,\ref{tab:description}.

\begin{table}[!t]
\centering
\renewcommand{\arraystretch}{1.8} 
\setlength{\tabcolsep}{8pt} 
\caption{Composition of our CoTA Dataset. We consider Google Gemini~\citep{gemini1.5} to build the reasoning ability in CoTA. Note that Multi-Speaker and Complex Audio datasets are manually synthesized, details of which can be referred to Sec.\,\ref{sec:synthetic}.}
\resizebox{\textwidth}{!}{
\begin{tabular}{llllccc} 
\toprule
\textbf{Category} & \textbf{Dataset Source} & \textbf{Main Skills Learning} & \textbf{Model Used} & \textbf{Quantity} & \textbf{Percentage} & \textbf{Synthetic} \\
\midrule
\multirow[c]{4}{*}{\textbf{Speech}} 
 & Multi-Speaker & Multi-speaker Speech QA & gemini-2.0-flash & 117.4k & 12.09\% & Yes \\ 
 & MELD~\citep{meld} & Speech Emotion QA & gemini-2.0-pro-exp & 29.2k & 3.01\% & No \\ 
 & CoVoST2~\citep{covost} & Speech-to-Text Translation & gemini-2.0-flash & 224.6k & 23.13\% & No \\ 
 & Mustard~\citep{mustard} & Sarcasm Detection& gemini-2.0-pro-exp & 1k & 0.10\% & No \\ 
\midrule
\multirow[c]{1}{*}{\textbf{Music}} 
 & MusicBench~\citep{mustango} & Music QA & gemini-2.0-flash & 137.1k & 14.12\% & No \\ 
\midrule
\multirow[c]{4}{*}{\textbf{Sound}} 
 & AudioSet~\citep{audioset} & Sound QA & gemini-2.0-flash & 315.2k & 32.46\% & No \\ 
 & Clotho~\citep{clotho} & Sound QA & gemini-2.0-pro-exp & 9.3k & 0.93\% & No \\ 
 & AudioCaps~\citep{audiocaps} & Sound QA & gemini-2.0-flash & 117.5k & 12.10\% & No \\ 
 & Complex Audio & Complex Audio QA & gemini-2.0-flash & 20k & 2.06\% & Yes \\ 
\bottomrule
\end{tabular}
}
\label{tab:composition}
\end{table}

\subsubsection{Conclusion and Next Steps}

In summary, we have introduced a systematic data generation pipeline that ensures the creation of high-quality, structured reasoning data for the Audio-Reasoner model. Our approach involves enriching raw audio data with detailed annotations, generating diverse questions, constructing explicit reasoning chains, and implementing a comprehensive validation framework. The following section provides an in-depth analysis of the final CoTA dataset and its reasoning capabilities, with a detailed statistical overview presented in Table\,\ref{tab:composition}.

\subsection{CoTA Dataset Analysis}

To evaluate the quality and reasoning efficacy of the CoTA dataset, we analyze its design from two key perspectives: (1) \textbf{comprehensive audio coverage}, ensuring broad representation across real-world and synthetic scenarios, and (2) \textbf{scalability of reasoning complexity}, which aligns task difficulty with structured inference patterns. Together, these aspects address critical gaps in audio-language pretraining.

\textbf{Comprehensive Audio Coverage}. CoTA integrates three audio domains---speech (38.33\%), music (14.12\%), and environmental sounds (47.55\%)---ensuring diverse and representative coverage of real-world auditory contexts. This multi-domain structure captures a wide spectrum of acoustic phenomena, ranging from conversational speech (\emph{e.g.}, speech-to-text translation tasks in \textit{CoVoST 2}) to intricate musical structures (\textit{MusicBench}) and fine-grained environmental sound analysis (\emph{e.g.}, \textit{AudioSet}'s rich descriptions of acoustic environments). 

A distinctive feature of CoTA is its hybrid synthetic-real composition, where synthetic data (Multi-Speaker and Complex Audio, 14.15\% of total samples) is strategically incorporated to enhance complex reasoning tasks, such as multi-step logical inference in \textit{Complex Audio}. Meanwhile, the majority of the dataset is derived from high-quality real-world sources (\emph{e.g.}, \textit{MELD} for emotion recognition). By unifying tasks across 10 distinct categories, spanning from fundamental classification to advanced tasks like translation and irony detection, CoTA facilitates a hierarchical learning process---an aspect largely absent in traditional datasets constrained to simple labeling tasks.



\begin{figure}[!t]
    \centering
    \includegraphics[width=1\textwidth]{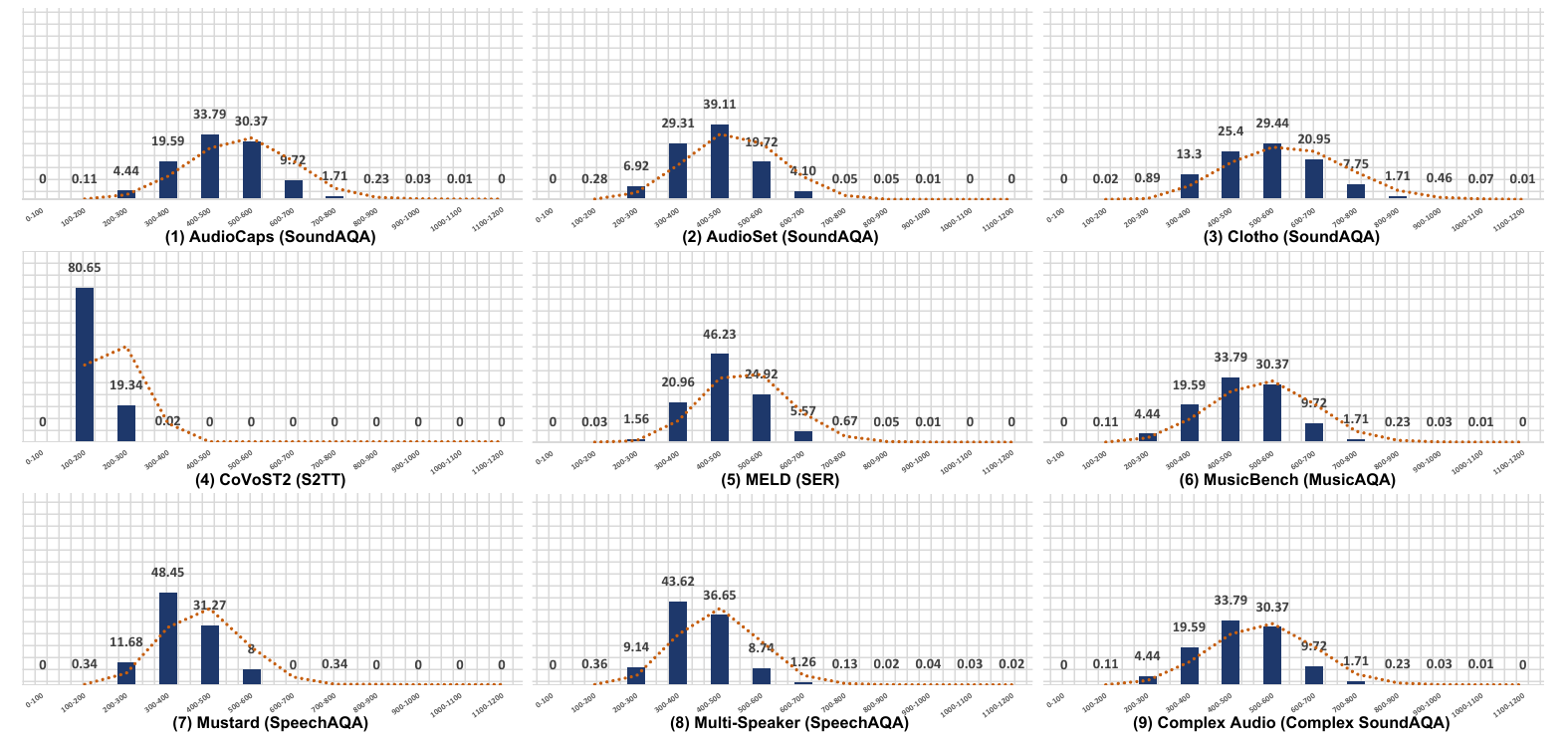} 
    \caption{The bar chart shows the data length distribution across nine CoTA sub-datasets, with intervals of 100 on the horizontal axis (0---1200) and proportions on the vertical axis (totaling 100\%). A moving average trend line is overlaid.}
    \label{fig:token_length_distribution}
\end{figure}

\textbf{Scalability of Reasoning Complexity}. 
The word count distribution in the CoTA dataset highlights the model’s capacity for long-chain reasoning. As shown in Figure\,\ref{fig:token_length_distribution}, most responses fall between 300 and 500 words, allowing for nuanced reasoning and detailed logical steps, particularly in audio and music question answering. This extended length supports transparency in reasoning, ensuring a thorough exploration of complex ideas.
For more demanding tasks, such as those in the Multi-Speaker dataset, responses can reach up to 1,500 words. This increase reflects the model’s ability to systematically break down intricate problems, demonstrating adaptive reasoning in scenarios requiring a deep understanding of multiple interacting elements.

Conversely, simpler tasks like S2TT elicit more concise responses, typically between 100 and 200 words. This brevity prevents unnecessary elaboration, maintaining clarity and efficiency when a straightforward answer suffices.
By adjusting response length based on task complexity, the model showcases its dynamic reasoning capabilities—--balancing depth and precision to effectively address a wide range of analytical challenges. Sec.\,\ref{sec:further dataset analysis} introduces more evidence about CoTA's scalability of reasoning complexity.

\section{Experiments}
\subsection{Experimental Setup}
\textbf{Training Details}. Our model, Audio-Reasoner, is built on Qwen2-Audio-Instruct~\citep{qwen2audio}, comprising a total of 8.4 billion parameters. The training process was conducted using the ms-swift framework~\citep{swift}, employing supervised fine-tuning with full-parameter tuning. We used a maximum learning rate of 1e-5 and trained the model for a single epoch across the entire CoTA dataset.

\begin{table}[!t]
    \centering
    \setlength{\tabcolsep}{4pt} 
    \caption{Summary of evaluation benchmarks for Audio-Reasoner.}
    \label{tab:qwen2-audio}
    \renewcommand{\arraystretch}{1.2} 
    \resizebox{\textwidth}{!}{
    \begin{tabular}{lllcl}
        \toprule
        \textbf{Dataset} &  \textbf{Split} & \textbf{Task} & \textbf{Metric} \\
        \midrule
        \multirow{3}{*}{MMAU-mini} 
            & Sound & Sound QA & ACC \\
        &  Speech & Speech QA&  ACC\\
        &  Music & Music QA &  ACC\\
        \midrule
        \multirow{9}{*}{AIR-Bench} 
        &  Chat-Sound & Sound QA& GPT-4 Eval \\
        &  Chat-Speech & Speech QA& GPT-4 Eval \\
        &  Chat-Music & Music QA & GPT-4 Eval \\
        &  Chat-MixedAudio & Complex Audio QA & GPT-4 Eval \\
        & Foundation-SoundAQA & Sound QA & ACC \\
        & Foundation-SER & Speech QA &  ACC\\
        & Foundation-SIC & Speaker Intent Classification&  ACC\\
        & Foundation-SNV & Speaker Number Verification&  ACC\\
        & Foundation-MusicAQA & Music QA& ACC \\
        \midrule
        CoVoST 2 & Test & Speech-to-Text Translation (S2TT)  & BLEU  \\

        \midrule
        MELD & Test & Speech Emotion Recognition (SER)  & ACC \\
       
        \bottomrule
    \end{tabular}
    }
    \label{tab:summary_evaluation}
\end{table}

\textbf{Evalutaion Metric}. Existing evaluation datasets in the audio domain primarily focus on simple classification tasks or BLEU\citep{bleu} score-based assessments of model responses. These approaches, however, fail to comprehensively reflect the reasoning capabilities of LALMs. To address this, our evaluation methodology incorporates multiple diverse benchmarks. 
First, we assessed the model's accuracy on closed-form questions using the MMAU-mini subset~\citep{mmau}. This subset was specifically chosen since the model's training data does not include multiple-choice questions. Additionally, we evaluated its real-world conversational abilities using the chat and foundation sections of AIR-Bench~\citep{air}. These benchmarks encompass various audio modalities, including sound, speech, and music.
Beyond these, we also tested the model's performance on traditional audio-related tasks such as speech-to-text translation (S2TT) on CoVoST 2~\citep{covost} and speech emotion recognition (SER) on MELD dataset~\citep{meld}. A comprehensive summary of the evaluation tasks and datasets is presented in Table\,\ref{tab:summary_evaluation}.

\textbf{Baselines:} We primarily select state-of-the-art large audio language models as the baselines for comparison. These include the closed-source models Gemini-1.5-pro~\citep{gemini1.5}, GPT-4o~\citep{gpt4o}, Qwen-audio-turbo~\citep{qwenaudio}, as well as the open-source models SALMONN~\citep{salmonn}, Qwen-Audio-Chat~\citep{qwenaudio}, and Qwen2-Audio-Instruct~\citep{qwen2audio} that also serves as the base model. Additionally, we compared cascade model approaches such as Whisper~\citep{radford2023robust} + GPT-4~\citep{gpt4} and a series of mainstream multimodal large language models.~\citep{LTU,LTUas,audioflamingo,gama,mullama,pandagpt,nextgpt,blsp,speechgpt}

\subsection{Main Results}
To evaluate the effectiveness of Audio-Reasoner, we compare its performance against both closed-source and open-source baselines on benchmark datasets of MMAU-mini and AIR-Bench chat/foundation, CoVoST 2 (zn/en subset) and MELD. The results in Tables\,\ref{table:performance_mini}, \,\ref{tab:performance_air_chat}, \,\ref{tab:performance_air_foundation}, \,\ref{tab:performance_covost2} and \,\ref{tab:performance_meld} clearly demonstrate that Audio-Reasoner significantly outperforms existing models, setting a new state-of-the-art in audio reasoning tasks.

\begin{table}[ht]
\caption{Performance comparison on MMAU-mini. The \textbf{\{so, mu, sp\}} indicates whether ``sound'', ``music'', and ``speech'' have been used in training.}
\centering
\setlength{\tabcolsep}{8pt}  
\begin{tabular}{l c c c c c c}
\toprule
\textbf{Model} & \textbf{Size} & \textbf{\{so, mu, sp\}} & \textbf{Sound} & \textbf{Music} & \textbf{Speech} & \textbf{Avg} \\
\midrule
\underline{\textit{\textbf{Closed-Source}}} & & & & & & \\
gpt4o + caption & - & - - - & 63.36 & 60.77 & 53.15 & 57.30 \\
gemini-1.5-pro & - & - - - & 56.75 & 49.40 & 58.55 & 54.90 \\
\midrule
\underline{\textit{\textbf{Open-Source}}} & & & & & & \\
LTU & 7B & Y  Y  N & 22.52 & 9.69 & 17.71 & 16.89 \\
LTU-AS & 7B & Y  Y  Y & 23.35 & 9.10 & 20.60 & 17.68 \\
Audio Flamingo - Chat & 2.2B & Y  Y  N & 23.42 & 15.26 & 11.41 & 16.69 \\
GAMA & 7B & Y  Y  N & 41.44 & 32.33 & 18.91 & 30.90 \\
GAMA-IT & 7B & Y  Y  N & 43.24 & 28.44 & 18.91 & 30.20 \\
MU-LLaMA & 7B & N  Y  N & 40.84 & 32.63 & 22.22 & 31.90 \\
SALMONN & 13B & Y  Y  Y & 41.00 & 34.80 & 25.50 & 33.70 \\
Qwen-audio-Chat & 8.4B & Y  Y  Y & 55.25 & 44.00 & 30.03 & 43.10 \\
Qwen2-Audio-Instruct & 8.4B & Y  Y  Y & 54.95 & 50.98 & 42.04 & 49.20 \\
\midrule
\underline{\textit{\textbf{Ours}}} & & & & & & \\
\textbf{Audio-Reasoner} & 8.4B & Y  Y  Y & \textbf{60.06} & \textbf{64.30} & \textbf{60.70} & \textbf{61.71 }\\
\bottomrule
\end{tabular}
\label{table:performance_mini}
\end{table}

\textbf{Performance on MMAU-mini}. 
MMAU-mini in Table\,\ref{table:performance_mini} assesses multimodal audio understanding across three major domains: sound, music, and speech. 
We first make a comparison with closed-source models.
Audio-Reasoner achieves the highest overall score (61.71\%) outperforming GPT-4o (57.30\%) and Gemini-1.5-Pro (54.90\%). 
Music reasoning shows the most significant improvement---Audio-Reasoner: 64.30\%, GPT-4o 60.77\% and Gemini-1.5-Pro: 49.40\%. 
This indicates superior musical structure comprehension, enabled by CoTA's diverse music-based tasks. Speech-based reasoning is also notably strong---Audio-Reasoner: 60.70\%, GPT-4o: 53.15\% and Gemini-1.5-Pro: 58.55\%.
This validates CoTA's impact in training models for context-dependent spoken language understanding. 
In comparison with open-source models, Audio-Reasoner surpasses all open-source models, with the next-best, Qwen2-Audio-Instruct, trailing by 12.51 percentage points (49.20\%). 
Across individual domains, Audio-Reasoner achieves 60.06\% in sound reasoning (beating Qwen2-Audio's 54.95\%), 64.30\% in music (outperforming Qwen2-Audio's 50.98\%), and 60.70\% in speech (exceeding Qwen2-Audio's 42.04\%).

\begin{table}[!t]
    \centering
    \caption{Performance comparison on AIR-Bench chat benchmark.}
    \begin{tabular}{lccccc}
        \toprule
        \textbf{Model} & \textbf{Sound} & \textbf{Music} & \textbf{Speech} & \textbf{Mixed Audio} & \textbf{Average} \\
        \midrule
        \underline{\textit{\textbf{Closed-Source}}} & & & & & \\
        Whisper+GPT4 & - & - & 7.54 & - & 7.54 \\
        Qwen-Audio-Turbo & 6.59 & 5.98 & 7.04 & 5.77 & 6.34 \\
        Gemini-1.5-pro & 5.49 & 5.06 & 6.97 & 5.27 & 5.70 \\
        \midrule
        \underline{\textit{\textbf{Open-Source}}}  & & & & & \\
        Macaw-LLM & 1.01 & 0.91 & 0.97 & 1.00 & 1.01 \\
        SpeechGPT & 0.95 & 0.95 & 1.57 & 1.14 & 1.15 \\
        Next-gpt & 4.76 & 4.18 & 3.86 & 2.92 & 4.13 \\
        Pandagpt & 5.46 & 5.06 & 3.58 & 2.93 & 4.25 \\
        BLSP & 5.55 & 5.08 & 6.17 & 4.52 & 5.33 \\
        Qwen-Audio & 6.95 & 5.52 & 6.47 & 5.38 & 6.08 \\
        SALMONN & 6.28 & 5.95 & 6.16 & 6.08 & 6.11 \\
        Qwen2-Audio-Instruct & 6.99 & 6.79 & 7.18 & \textbf{6.77} & 6.93 \\
        \midrule
        \underline{\textit{\textbf{Ours}}}  & & & & & \\
        \textbf{Audio-Reasoner} & \textbf{7.68} & \textbf{8.05} & \textbf{8.19} & 6.65 & \textbf{7.94} \\
        \bottomrule
    \end{tabular}
    \label{tab:performance_air_chat}
\end{table}

\textbf{Performance on AIR-Bench chat}. 
\textit{(1) chat benchmark}.
AIR-Bench chat in Table\,\ref{tab:performance_air_chat} evaluates contextutal and conversational reasoning across four audio types: sound, music, speech, and mixed audio.
Regarding closed-source models, Audio-Reasoner achieves the highest overall score (7.94), outperforming Gemini-1.5-Pro (5.70) and Whisper+GPT-4 (7.54). It shows the most significant improvements in music (8.05) and speech (8.19). Additionally, its mixed audio reasoning score (6.65) demonstrates proficiency in handling multi-source audio tasks.
As for comparison with open-source models, Audio-Reasoner sets a new benchmark, surpassing Qwen2-Audio (6.93) by 1.01 points. Across domains, it achieves 7.68 in sound (beating Qwen2-Audio's 6.99), 8.05 in music (exceeding Qwen2-Audio's 6.79), and 8.19 in speech (outperforming Qwen-2-Audio's 7.18)
, showcasing balanced expertise.

\begin{table}[hb]
    \caption{Performance comparison on AIR-Bench foundation benchmark.}
    \centering
    \setlength{\tabcolsep}{8pt}
    \begin{tabular}{l c c c c c c c}
        \toprule
        \multirow{2}{*}{\textbf{Model}} & \multicolumn{1}{c}{\textbf{AIR-Sound}} & \multicolumn{1}{c}{\textbf{AIR-Music}} & \multicolumn{3}{c}{\textbf{AIR-Speech}} & \multicolumn{1}{c}{\textbf{Average}} \\
        \cmidrule(lr){2-2} \cmidrule(lr){3-3} \cmidrule(lr){4-6} \cmidrule(lr){7-7}
         & \textbf{SoundAQA}  & \textbf{MusicAQA} & \textbf{SER} & \textbf{SIC} & \textbf{SNV} & \textbf{} \\
        \midrule
        \underline{\textit{\textbf{Closed-Source}}} & & & & & & \\
        whisper+GPT4 & - & -  & 59.5 & 87.7 & 30.0 &  59.1 \\
        Qwen-Audio-Turbo & 62.8 & 62.5 & 60.0 & 56.4 & 54.3 &  59.2\\ 
         \midrule
        \underline{\textit{\textbf{Open-Source}}} & & & & & & \\
        NEXT-GPT & 18.8 & 47.1 & 25.7 & 25.6 & 25.4 & 28.5 \\
        SpeechGPT & 33.9 & 31.3 & 37.6 & 45.8 & 32.6 &  36.2\\
        BLSP & 36.1  & 31.0 & 27.4 & 46.6 & 28.1 &  33.8\\
        PandaGPT & 48.7 & 50.7 & 26.0 & 28.5 & 43.2 &  39.4\\
        SALMONN & 28.4& 54.6 & 29.9 & 36.7 & 34.3 &  36.8\\
        Qwen-Audio-Chat & 64.6 & 48.2 & 43.2 & 77.8 & 35.3 & 53.8 \\
         \midrule
        \underline{\textit{\textbf{Ours}}} & & & & & & \\
        \textbf{Audio-Reasoner} & \textbf{65.7} &\textbf{ 55.2} &\textbf{ 60.5} &\textbf{ 88.1 }& \textbf{56.3} & \textbf{65.2}\\
        \bottomrule
    \end{tabular}
        \label{tab:performance_air_foundation}
\end{table}

\textit{(2) foundation benchmark}. 
AIR-Bench foundation in Table\,\ref{tab:performance_air_foundation} evaluates fundamental audio understanding across three primary categories: sound, music, and speech, with speech further divided into three subdomains: Speech Emotion Recognition (SER), Speaker Identification and Classification (SIC), and Speech Number Variation (SNV).
Audio-Reasoner achieves the highest overall score (65.2), outperforming both closed-source and open-source baselines. Compared to the strongest closed-source model, Qwen-Audio-Turbo (59.2), Audio-Reasoner leads by 6.0 points, demonstrating superior reasoning across all audio domains.
With the sound category, Audio-Reasoner attains 65.7, surpassing Qwen-Audio-Chat (64.6) and Qwen-Audio-Turbo (62.8), highlighting its strong ability in environmental and non-speech audio understanding.
For music reasoning, Audio-Reasoner achieves 55.2, significantly outperforming Qwen-Audio-Turbo (48.2) and all open-source baselines, indicating better comprehension of musical structures and attributes.
Regarding speech reasoning, Audio-Reasoner sets new state-of-the-art results across all subdomains. It attains 60.5 in SER (\emph{v.s.} Qwen-Audio-Turbo's 60.0), 88.1 in SIC (surpassing Whisper+GPT-4's 87.7), and 56.3 in SNV (exceeding Qwen-Audio-Turbo's 54.3). The substantial lead in SIC showcases its exceptional speaker recognition capability, benefiting from CoTA’s step-by-step reasoning process.

\begin{table}[ht]
    \centering
    \caption{Performance comparison of the speech-to-text translation (S2TT) task on CoVoST 2 dataset. We consider the mutual conversion between Chinese and English as training and evaluation data.}
    \small
    \setlength{\tabcolsep}{1pt}
    \begin{tabular}{lcccccccccccc}
        \toprule
        \multirow{2}{*}{\textbf{Model}} & \multicolumn{5}{c}{\textbf{EN-ZN}} & \multicolumn{5}{c}{\textbf{ZN-EN}} & \multirow{2}{*}{\textbf{Avg}} \\
        \cmidrule(lr){2-6} \cmidrule(lr){7-11}
        & BLEU1 & BLEU2 & BLEU3 & BLEU4 & Avg & BLEU1 & BLEU2 & BLEU3 & BLEU4 &  Avg & \\
        \midrule
        
        \underline{\textit{\textbf{Closed-Source}}} & & & & & & \\
       Gemini-1.5-pro & 68.25 & 49.12 & 37.81 & 29.79 & 46.24  & 51.83 & 26.61 & 16.27 & 10.88 & 26.39 & 36.32\\
      
        \underline{\textit{\textbf{Open-Source}}} & & & & & & \\
        Qwen2-Audio-Instruct & 58.63 & 39.55 & 28.71  & 21.40 & 37.07  & 48.52 & 24.31 & 14.65 & 9.24 & 24.18 & 30.63\\
         \underline{\textit{\textbf{Ours}}} & & & & & & \\
        Audio-Reasoner & \textbf{72.89} & \textbf{54.17} & \textbf{42.46} & \textbf{33.95} & \textbf{50.87} & \textbf{56.50} & \textbf{29.99} & \textbf{18.37}  & \textbf{11.62} & \textbf{29.13} & \textbf{40.00}\\        
        \bottomrule
    \end{tabular}
    \label{tab:performance_covost2}
\end{table}

\textbf{Performance on CoVoST 2}. 
The CoVoST 2 dataset in Table\,\ref{tab:performance_covost2} evaluates speech-to-text translation, a fundamental task in cross-lingual speech understanding. Audio-Reasoner demonstrates the strengths of Audio-Reasoner in speech-to-text translation across both English-to-Chinese (EN-ZN) and Chinese-to-English (ZN-EN) tasks.

For \textit{EN-ZN translation}, Audio-Reasoner outperforms both closed-source Gemini-1.5-pro and open-source Qwen2-Audio-Instruct. With an average BLEU score of 50.87, it surpasses Gemini-1.5-pro's score of 46.24 by 4.63 points and Qwen2-Audio-Instruct's 37.07 by a significant 13.80 points. Audio-Reasoner's BLEU-4 score of 33.95 highlights its ability to generate fluent, high-quality translations, even for more complex sentence structures.
In the \textit{ZN-EN translation} task, Audio-Reasoner continues to show superiority with an average BLEU score of 29.13, outperforming Gemini-1.5-pro (with a score of 26.39) by 2.74 points and Qwen2-Audio-Instruct (with a score of 24.18) by 4.95 points. Its BLEU-4 score of 11.62 reflects an enhanced ability to produce coherent and accurate translations, particularly in more challenging, longer sentences.

These results demonstrate that Audio-Reasoner excels in both capturing cross-lingual semantic alignment, consistently outperforming existing models in speech-to-text translation tasks.

\begin{table}[ht]
    \centering
    \caption{Performance comparison of the speech emotion recognition (SER) task on MELD dataset.}
    \begin{tabular}{lc}
        \hline
        Model & Unweighted\_ACC \\
        \hline
        EMO-box & 31.5 \\
        SALMONN & 39.2 \\
        Qwen2-Audio-Instruct & 49.9 \\
        Audio-Reasoner & 53.9 \\
        \hline
    \end{tabular}
    \label{tab:performance_meld}
\end{table}

\textbf{Performance on MELD}.
The MELD dataset in Table\,\ref{tab:performance_meld} evaluates speech emotion recognition, a crucial task for understanding affective states in spoken language. Audio-Reasoner achieves the highest unweighted accuracy of 53.9, outperforming all baseline models, including the previous state-of-the-art Qwen2-Audio-Instruct (49.9) by 4.0 percentage points.
The results further validate that Audio-Reasoner excels in emotion perception and contextual audio understanding, solidifying its effectiveness in real-world speech emotion recognition tasks.

\textbf{Why Audio-Reasoner Excels}.
While open-source models typically lag behind closed-source alternatives, Audio-Reasoner narrows this gap significantly. It outperforms Gemini-1.5-Pro in multiple categories and achieves competitive results against GPT-4o.
A key factor driving Audio-Reasoner's state-of-the-art performance is the CoTA dataset, which provides an unparalleled foundation for comprehensive, structured, and context-aware audio reasoning. The dataset contributes to Audio-Reasoner's success in two critical ways:
\textit{(1) CoTA's comprehensive audio reasoning data}.
Unlike models trained on fragmented or task-specific datasets, Audio-Reasoner benefits from CoTA's extensive and balanced coverage of sound, speech, and music-based reasoning. This ensures robust adaptation across diverse audio tasks, enhancing its ability to generalize across multimodal benchmarks.
\textit{(2) CoTA's strong logical and contextual inference}. 
CoTA’s emphasis on structured reasoning and contextual awareness enables Audio-Reasoner to outperform existing models in long-form reasoning (MMAU-mini) and conversational audio understanding (AIR-Bench chat). By leveraging deep logical inference and nuanced context retention, it surpasses prior approaches in handling complex speech interactions and multimodal reasoning.

These strengths position Audio-Reasoner as a breakthrough in open-source audio intelligence, setting a new benchmark for multimodal reasoning in speech comprehension, translation, and real-world audio interpretation.

\begin{figure}[!t]
    \centering
    \includegraphics[width=1.0\linewidth]{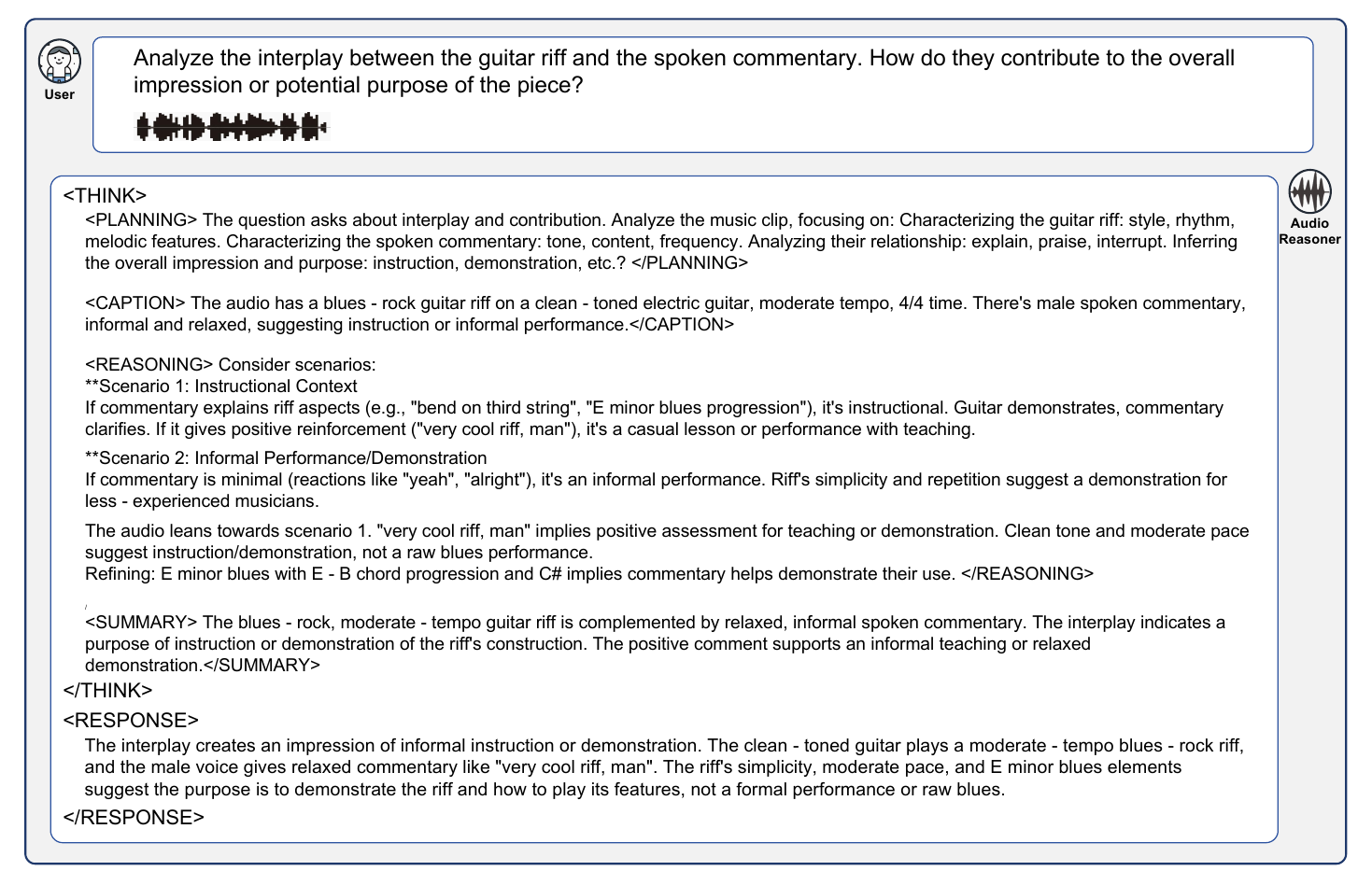}
    \caption{Example of Audio-Reasoner answering music-based question.}
    \label{fig:case_study}
\end{figure}

\subsection{Case Study}
Here we show a case in Figure\,\ref{fig:case_study}, demonstrating the audio-based reasoning capability of Audio-Reasoner. The system analyzes the interplay between a guitar riff and a spoken commentary, systematically breaking down their characteristics and relationship. It identifies key musical features, assesses the commentary's tone and intent, and infers the overall purpose of the piece. By considering different scenarios, Audio-Reasoner determines that the interplay suggests an informal instructional or demonstrative context rather than a formal performance. 
This example highlights the model's ability to extract meaningful insights from audio, combining musical analysis with contextual interpretation.

For more remarkable examples, please refer to Sec.\,\ref{sec:more_cases}.

\section{Conclusion}
In this work, we introduced Audio-Reasoner, a large audio language model (LALM) designed to advance deep reasoning in audio-based tasks. By leveraging inference scaling and structured chain-of-thought (CoT) reasoning, we demonstrated significant performance improvements across key benchmarks. Central to our approach is CoTA, a large-scale, high-quality dataset containing around 1.2 million structured reasoning samples, which we generated through a systematic pipeline of annotation refinement, question synthesis, and CoT generation.

Our experimental results highlights the effectiveness of structured reasoning in the audio domain, achieving state-of-the-art performance on MMAU-mini (+25.04\%), CoVoST 2 (+8.31\%), and MELD (+8.01\%). These findings underscore the critical role of reasoning-rich datasets and inference scaling in multimodal learning, particularly for audio-based tasks where existing models struggle with complex reasoning.

Moving forward, we believe that structured CoT methodologies will play a crucial role in enhancing the reasoning capabilities of LALMs. Future research directions include extending Audio-Reasoner's capabilities to multi-turn dialogue reasoning, cross-modal knowledge integration, and real-world applications in speech-driven AI systems. By publicly releasing our model, dataset, and code, we aim to foster further innovation and collaboration in the field of audio reasoning, paving the way for more advanced and interpretable LALMs.

\bibliographystyle{plainnat} 
\bibliography{ref}

\newpage

\appendix

\section{Prompt Details}
\raggedbottom
A universally applicable method for writing prompts involves three key components: a clear task definition, a structured example, and a precise format specification. Our prompt adheres to this methodology by first defining the task explicitly, outlining the need for detailed audio descriptions and progressively challenging questions. It then provides a structured example that demonstrates the expected output format, ensuring clarity and minimizing ambiguity. Lastly, it specifies the exact formatting rules using delimiters such as <caption>...</caption> and <question1>...</question1>, ensuring consistency in responses. This approach guarantees efficiency by eliminating interpretative variance, allowing for precise and reproducible outputs. When drafting this prompt, we adhered to a structured approach to maximize clarity and effectiveness. The first-person perspective is used to emphasize our direct involvement in designing the task, ensuring the reader understands the rationale behind each structural choice. The structure follows a logical progression: we begin by introducing the general method, transition into an explanation of how our prompt aligns with this method, and conclude by justifying the approach’s efficiency. By maintaining an academic tone, we reinforce the credibility and rigor of our prompt-writing methodology. We list some of the prompts used for tasks at different stages following below.

\subsection{Prompt of Stage 1 when Processing Data (Sample from AudioSet)\label{sec:prompt_stage1}}

\begin{tcolorbox}[width=1.0\textwidth]
    We are annotating some audio and designing some questions. You are an excellent audio analyst. Next, you will receive an audio and one absolutely correct but simple description. Your task is to first generate a more detailed, in-depth and absolutely correct new description based on the given descriptions. Then, use this description to generate three open-ended or single-choice questions with four options along with their answers. Please separate different parts using 
    <caption>...</caption> <question1><question>...</question> <answer>...</answer></question1> <question2> <question>...</question> <answer>...</answer></question2>......\\
    
    Here is a sample. Please strictly follow the format in the sample. 
    <caption>The audio presents a sustained, high-frequency static noise, characteristic of a detuned or malfunctioning electronic device, likely a television or radio...</caption><question1><question>Describe the characteristics of the static noise in the audio, and how these characteristics change over time.</question><answer>...</answer></question1><question2> <question>What...?</question> <answer>...</answer> </question2> <question3><question>What...?</question> <answer>...</answer> </question3>\\
    
    Here is the original description: \textbf{*** label here ***}. 
    
    Here is the audio.
    
    Please design three questions that gradually become more challenging, starting from basic factual questions, but don't deviate from the content of the audio itself. If it's a single-choice question, please give four options like (a) one, (b) two, .... and the answer should be analyzed and end with a format like ``the answer is B.'' 
\end{tcolorbox}

\subsection{Prompt of Stage 2 when Processing Data (Sample from AudioSet)\label{sec:prompt_stage2}}
\begin{tcolorbox}[width=1.0\textwidth]
    We are now designing a system to generate structured audio-based chain-of-thought reasoning data. You will receive an audio clip, its textual description, as well as a question and its answer. Your task is to explore in more detail the thinking process from the question to the answer. Your chain of thought will be divided into two main parts, the <think> and <response> sections. In the <think> section, you need to go through four steps: planning, captioning, reasoning, and summarizing. The <think> section is invisible to the user. Therefore, in the <response> section, you need to base on all the reasoning processes and results in the <think> section and provide a final reply based on the question. Finally, your reply should strictly follow the following format: <THINK><PLANNING> (In this part, analyze the user's needs and how to complete the task. If the problem is complex, it should be done step by step) </PLANNING><CAPTION> (In this part, conduct an overall analysis of the given audio input content, try to find all the parts related to the question, describe them, and ensure it is completely correct.) </CAPTION><REASONING> (Start reasoning towards the final answer here) </REASONING><SUMMARY> (Draw appropriate results based on the reasoning part here) </SUMMARY></THINK><RESPONSE> Give the final answer here referring to the <THINK> part </RESPONSE> Please strictly follow the format of the sample.\\

    Sample:    \\
    <THINK> \\
    <PLANNING>\\
    The user wants to understand the dynamic changes within the provided audio clip ......\\</PLANNING>\\
    <CAPTION>\\
    The audio clip predominantly features static noise. ...... similar to that of a detuned television or a device failing to receive a signal.\\</CAPTION>\\
    <REASONING>\\
    1.  Identify changes in Intensity (Volume): The audio's static noise does not remain at a constant volume. There are noticeable increases and decreases in loudness throughout the clip. ......  is struggling to maintain a consistent output, adding to the impression of something malfunctioning or broken.\\</REASONING>\\
    <SUMMARY>\\
    The static noise in the audio is highly dynamic. ...... leading to a sense of disorder and instability.\\</SUMMARY>\\
    </THINK>\\
    <RESPONSE>\\
    The audio presents a static noise, ......The overall effect is one of energetic chaos, preventing any possibility of calm or predictability.\\</RESPONSE>\\
        
    Note that you have both the question and the answer because it is necessary to ensure the correctness of the chain of thought. However, in your response, you can only refer to the content of the question and the audio, and lead to the answer. You must absolutely not assume that you already know the answer. Please provide a detailed and flexible response with high-quality logic in both the caption and reasoning sections. If the reasoning part requires complex logic, you can even propose several different approaches and try them one by one.
    
    Here is the original description: \textbf{*** caption here ***}. 
    
    The question is: \textbf{*** question here ***}. 
    
    The answer you can refer to : \textbf{*** answer here ***}. 
    
    Again, don't mention that you have the answer and the description because they are only here to help you to design the chain of thought but should not exist in the real-world scenario, either in the think or response sections. 
\end{tcolorbox}

\subsection{Prompt of Stage 3 when Processing Data (Sample from AudioSet)\label{sec:prompt_stage3}}
\begin{tcolorbox}[width=1.0\textwidth]
    We are data reviewers. Next, you will receive an audio clip, along with its description, questions, answers, and most importantly, the thought process for solving the problems. Please determine and analyze whether all of these elements are completely correct, especially check if there are any hallucinations in the thought process. Return <True> if there are no issues, and <False> if there are errors in the data.
    
    Here is the description of the audio: 
    \textbf{*** caption here ***}. 
    
    Here is the question: 
    \textbf{*** question here ***}. 
    
    Here is the answer: 
    \textbf{*** answer here ***}. 
    
    And here is the thought process: \textbf{*** COT process here ***}. 
    
    Please conduct a thorough judgment and analysis and provide the result in the specified format.
\end{tcolorbox}

\section{Synthetic Data Generation Pipeline\label{sec:synthetic}}
\subsection{Synthetic Data Introduction\label{sec:synthetic_intro}}
\textbf{Multi-Speaker Dataset}: To enhance the model's ability to comprehend complex, multi-turn conversations among multiple speakers, we constructed the Multi-Speaker dataset using text-to-speech (TTS) technology. The dataset generation process consists of three steps: (1) \textbf{Conversation Text Generation}: We utilized commericial models to generate diverse multi-speaker conversation texts covering a wide range of scenarios. 
(2) \textbf{Speech Synthesis}: Leveraging all available timbres from LibriSpeech~\citep{librispeech} as prompts, we employed the CosyVoice2~\citep{du2024cosyvoice} framework to synthesize high-quality speech samples. 
(3) \textbf{Dataset Assembly}: The synthesized speech samples, feauring distinct timbres, were carefully combined to create a rich and diverse multi-speaker conversation dataset.

This approach ensures that the dataset effectively trains the model to navigate intricate speaker interactions in natural conversations.

\textbf{Complex Audio Dataset}: To strengthen the model's ability to process intricate audio compositions, we developed the Complex Audio dataset. The audio materials are sourced from AudioSet, AudioCaps, and our internal material library. The dataset consists of two primary types of audio data:
(1) \textbf{Temporally Complex Audio}: Short audio clips were sequentially arranged, requiring the model to perform tasks such as sorting or counting based on temporal relationships.
We sequentially arranged different short audio clips and required the model to perform sorting or counting tasks. 
(2) \textbf{Composite Complex Audio}: Multiple long-form audio tracks---such as engine sounds and musical instruments--were blended. The model was then tasked with extracting specific target sounds from the composite audio, enhancing its ability to analyze overlapping audio sources.

This dataset ensures that the model is well-equipped to handle real-world audio challenges, such as distinguishing between multiple concurrent sounds and identifying temporal patterns.

To better understand what Multi-Speaker dataset and Complex Audio dataset are respectively tasked with, we provide some examples in Sec.\,\ref{sec:synthetic_samples_1}, Sec.\,\ref{sec:synthetic_samples_2}.

\subsection{Sample from Complex Audio Dataset\label{sec:synthetic_samples_1}}
\begin{tcolorbox}[width=1.0\textwidth]
<THINK>

<PLANNING>

The question is about identifying the sounds in the audio and arranging them in the correct order. I will carefully listen to the audio multiple times, noting down the sounds as I recognize them. Then, I will compare the identified sound sequence with the provided answer options to choose the one that matches best.\\</PLANNING>

<CAPTION>Upon initial listening, I can discern sounds that seem like a shattering noise, an animal sound which I think sounds like a cow, a cat sound, and what sounds like some mechanical tools working, perhaps a chainsaw. I shall listen again to clarify the sounds and their order.\\</CAPTION>

<REASONING>1. After relistening, the first sound is definitely glass breaking.\\2. The second sound is definitely a cow mooing.\\3. The third sound is a cat meowing.\\4. The fourth sound is a chainsaw.\\5. I compare this sequence (glass break, cow, cat, chainsaw) with the options provided.\\6. Option (a) matches the order I identified.\\</REASONING>

<SUMMARY>Based on the detailed listening and analysis, the correct order of sounds in the audio is: glass break, cow, cat, chainsaw. This matches option (a).\\</SUMMARY>

</THINK>

<RESPONSE>

The answer is a.

</RESPONSE>
\end{tcolorbox}

\subsection{Sample from Multi-Speaker Dataset\label{sec:synthetic_samples_2}}

\begin{tcolorbox}[width=1.0\textwidth]
<THINK>\\
<PLANNING>\\
The user asks me to identify the number of speakers in the audio clip. To do this, I will listen carefully and try to distinguish unique voices. The conversation is between two hikers, I need to confirm this by identifying their individual lines and vocal characteristics. Finally, I will provide the answer based on the number of identified speakers.\\
</PLANNING>\\
<CAPTION>\\
The audio presents a conversation between two individuals hiking.\\
Speaker 1: This person starts by commenting on the intensity of the climb and expresses doubt about continuing without a break. Later, after a reminder about the ranger's warning, this speaker agrees to turn back. The vocal tone suggests exertion and a desire to reach the summit, eventually giving way to agreement. Phrases spoken: ``Wow, this climb is intense. I am not sure I can go much further without a serious break.'', ``Turn back. But we're so close to the summit. I really want to make it, you know.'', ``Unpredictable mountain weather, okay. Okay, you are right.''\\
Speaker 2: This person responds to the first speaker's concerns, suggesting they turn back due to the darkening sky and diminishing daylight. They emphasize safety and remind the other speaker about the ranger's warning. The vocal tone conveys concern and prioritizes safety. Phrases spoken: ``I know right. I am starting to think we should turn back. That sky looks awfully dark, and we're losing daylight quickly.'', ``I hear you, but safety first. Remember what the ranger said about the unpredictable mountain weather.'', ``So, turn back.''\\
Based on the clear distinction in vocal tones and the back-and-forth nature of the dialogue, I can confirm that there are two distinct speakers in this audio clip.\\
</CAPTION>\\
<REASONING>\\
Initial Assessment: The audio starts with two distinct voices in a hiking - related dialogue. The voices are clearly distinguishable from the get - go, with one having a deeper timbre and the other a higher - pitched one.\\
Dialogue Analysis: A close look at the content shows a clear difference in the speakers' stances. One is eager to press on to the summit, saying, ``We're almost there. Let's keep going.'' The other is more cautious, citing the darkening sky and fading daylight, stating, ``The weather looks bad. We should turn back.''\\
Vocal Tone Differentiation: Each speaker's vocal tone stays consistent. The eager speaker has an energetic and determined tone, with a rising pitch when advocating for continuing. The cautious speaker has a measured, worried tone, with a lower pitch when warning of danger.\\
Confirming Speaker Count: A review of the audio shows no extra voices or interruptions. The conversation moves smoothly between the two speakers, indicating only two are present.\\
Final Answer: Based on this analysis, there are two individuals in the conversation. \\Thus, the answer is B.\\
</REASONING>\\
<SUMMARY>\\
Based on the dialogue and vocal tone analysis, there are two distinct speakers in the audio. The answer is B.\\
</SUMMARY>\\
</THINK>\\
<RESPONSE>\\
Based on the dialogue and distinct vocal tones present in the audio, I have identified two individuals engaged in the conversation. Therefore, the answer is B.\\
</RESPONSE>
\end{tcolorbox}

\newpage
\section{Further Dataset Analysis\label{sec:further dataset analysis}}

\begin{figure}[ht]
    \centering
    \includegraphics[width=1\textwidth]{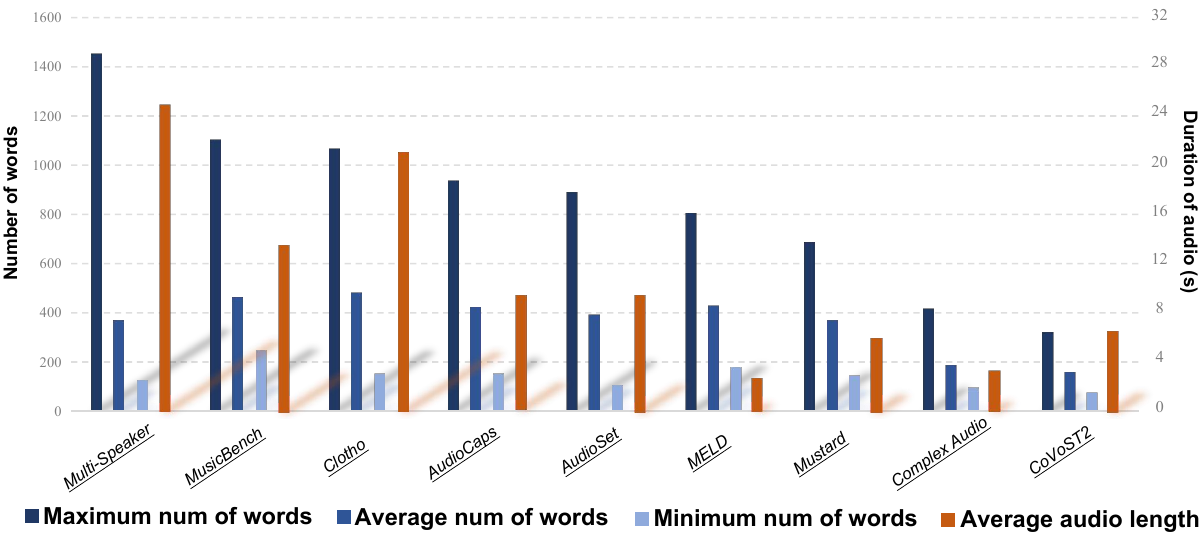} 
    \caption{Detailed information of the CoTA dataset: The maximum, minimum, and average number of words in the answers, as well as the average length of the audio.}
    \label{fig:text_and_audio_length}
\end{figure}

CoTA's reasoning adaptability is also evident in its distribution of token lengths and audio duration across tasks in Figure\,\ref{fig:text_and_audio_length} . The dataset accommodates a broad range of reasoning complexities, with token lengths spanning 327 (CoVoST2) to 1,454 (Multi-Speaker), ensuring coverage of both concise and highly intricate reasoning processes.
Notably, tasks requiring deep logical inference, such as complex audio, exhibit a well-balanced token distribution (max = 423, avg = 192.96), supporting structured multi-step reasoning without unnecessary redundancy.

Further, the dataset's average token lengths (164.48–481.57) align with task difficulty: longer reasoning chains characterize tasks such as sound description (AudioSet: 395.26) and music understanding (MusicBench: 463.89). Meanwhile, CoTA ensures practical generalization by maintaining audio durations between 2.85s and 26.34s, where shorter clips (\emph{e.g.}, MELD: 2.84s) support concise context-dependent reasoning, while extended sequences (Multi-Speaker: 26.34s) enable complex multi-turn inference. This systematic variation in reasoning depth and audio granularity ensures adaptability across diverse tasks, addressing the limitation of one-size-fits-all reasoning chains in existing audio datasets.

\newpage
\section{More Case Studies\label{sec:more_cases}}

\begin{figure}[ht]
    \centering
    \includegraphics[width=1\textwidth]{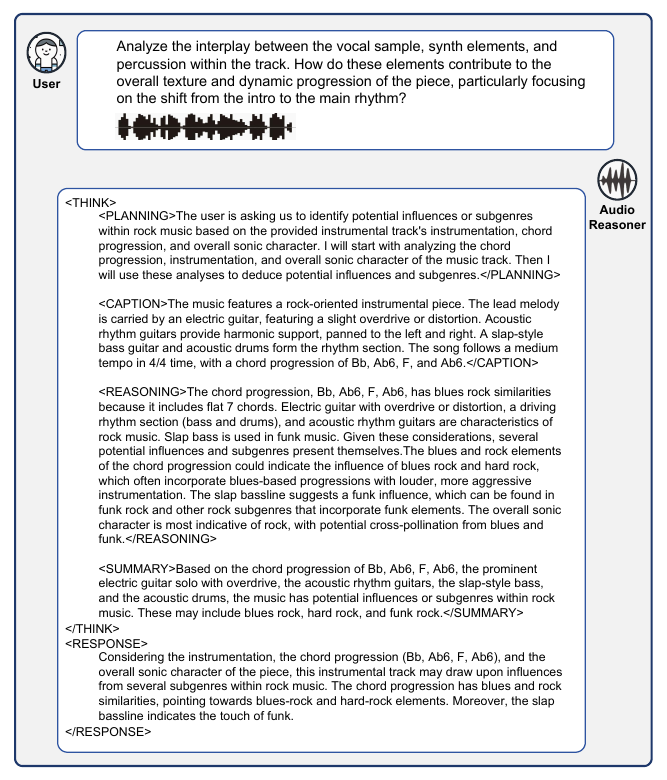} 
    \caption{Example of Audio-Reasoner answering music-based question.}
    \label{fig:case2}
\end{figure}

\begin{figure}[ht]
    \centering
    \includegraphics[width=1\textwidth]{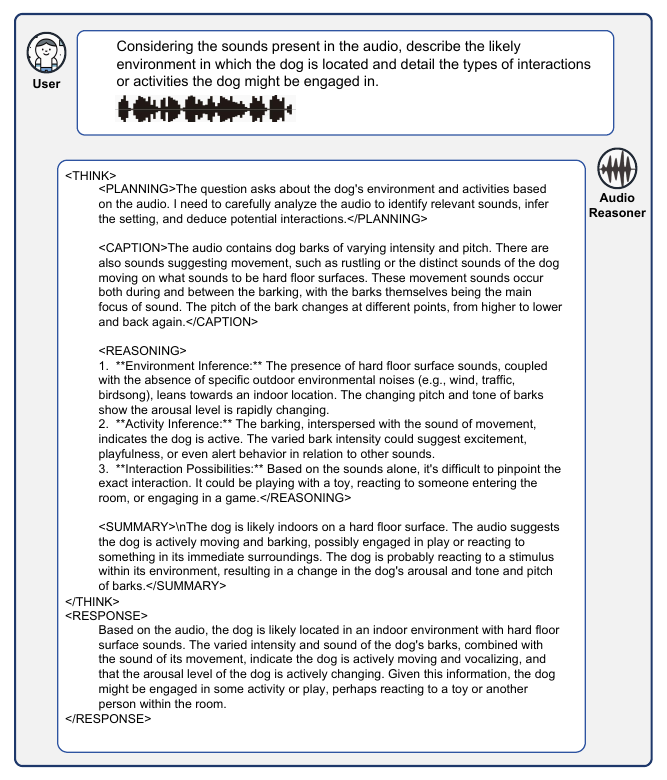} 
    \caption{Example of Audio-Reasoner answering sound-based question.}
    \label{fig:case3}
\end{figure}

\begin{figure}[ht]
    \centering
    \includegraphics[width=1\textwidth]{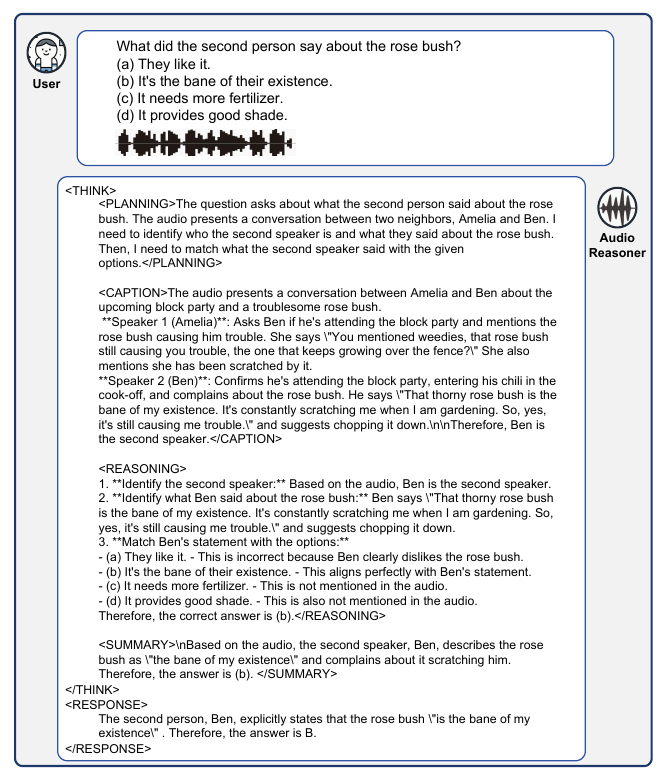} 
    \caption{Example of Audio-Reasoner answering multi-speaker speech-based question.}
    \label{fig:case4}
\end{figure}

\begin{figure}[ht]
    \centering
    \includegraphics[width=1\textwidth]{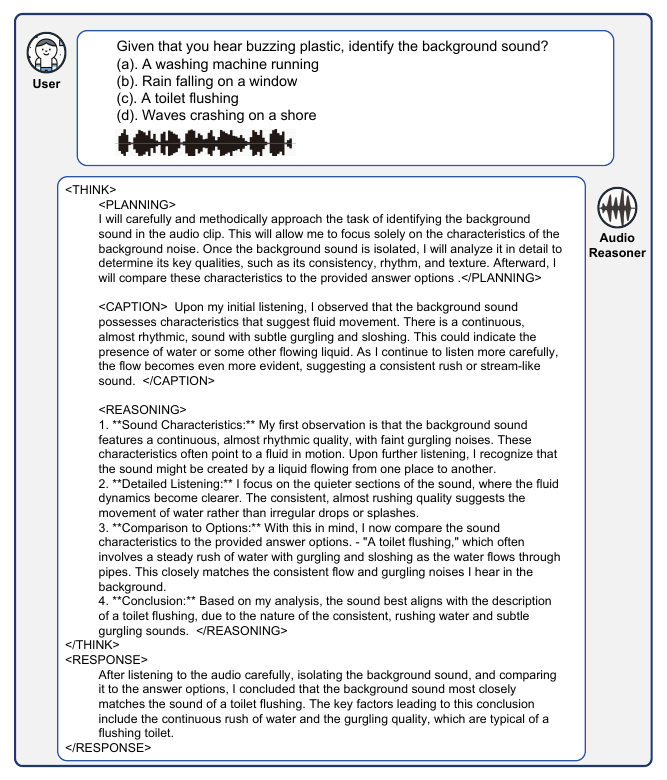} 
    \caption{Example of Audio-Reasoner answering complex-audio-based question.}
    \label{fig:case5}
\end{figure}
\end{document}